\magnification = \magstep1
\baselineskip=16pt
\def\ni{\medskip\noindent}
 
\def\abst#1{\medskip{\baselineskip=14pt
{\narrower\narrower\parindent = 0pt #1 \par}}} 
\def\qquote#1{\medskip{\baselineskip=14pt
{\narrower\narrower\parindent = 0pt #1 \par}}} 

\def\>{\rangle}
\def\<{\langle}
\def\k#1{|#1\>}

\def \ip#1#2{\< #1 | #2 \>}

\newcount\ftnumber
\def\ft#1{\global\advance\ftnumber by 1
          {\baselineskip 12pt    
           \footnote{$^{\the\ftnumber}$}{#1}}}

\newcount\eqnumber

\def\eq(#1){
    \ifx\DRAFT\undefined\def\DRAFT{0}\fi	
    \global\advance\eqnumber by 1%
    \expandafter\xdef\csname !#1\endcsname{\the\eqnumber}%
    \ifnum\number\DRAFT>0%
	\setbox0=\hbox{#1}%
	\wd0=0pt%
	\eqno({\offinterlineskip
	  \vtop{\hbox{\the\eqnumber}\vskip1.5pt\box0}})%
    \else%
	\eqno(\the\eqnumber)%
    \fi%
}
\def\(#1){(\csname !#1\endcsname)}

\centerline{{\bf Nonlocal character of quantum theory?}}
\bigskip
\centerline{N. David Mermin}
\centerline{Laboratory of Atomic and Solid State Physics}
\centerline{Cornell University, Ithaca, NY 14853-2501}
\medskip
\abst{In a recent article under the above title (but without the
question mark) Henry Stapp presented arguments which lead him to
conclude that under suitable conditions ``the truth of a statement
that refers only to phenomena confined to an earlier time'' must
``depend on which measurement an experimenter freely chooses to
perform at a later time.'' I point out that the reasoning leading to
this conclusion relies on an essential ambiguity regarding the meaning
of the expression ``statement that refers only to phenomena confined to
an earlier time'' when such a statement contains counterfactual
conditionals.  As a result the argumentation does not justify the
conclusion that there can be frames of reference in which future
choices can affect present facts.  But it does provide an instructive
and interestingly different opportunity to illustrate a
central point of Bohr's reply to Einstein, Podolsky, and Rosen.} 
\bigskip
\centerline{{\bf I.  Introduction.}}
\medskip Henry Stapp\ft{Henry P.~Stapp, ``Nonlocal character of quantum
theory,'' Am.~J.~Phys.~{\bf 65}, 300-304 (1997).} has subjected the
strange quantum correlations discovered by Lucien Hardy\ft{Lucien
Hardy, ``Quantum mechanics, local realistic theories, and Lorentz
invariant realistic theories,'' Phys.~Rev.~Lett.~{\bf 68}, 2981-2984
(1992).} to a cunning logical analysis.  In a series of fourteen
propositions, each of which appeals only to the predictions of quantum
mechanics, formal logical transformations, or one of three locality
conditions, Stapp reaches a contradiction leading him to reject at
least one of the locality conditions and conclude that ``the general
causal principle that the present facts are independent of free
choices to be made at later times'' cannot hold in all Lorentz
frames.  

Many of Stapp's propositions contain counterfactual conditionals ---
statements about the outcome of experiments that might have been
performed but were not.  Such constructions, Stapp emphasizes, are
permissible provided they do nothing more than state the consequences,
for experiments that might have been performed but were not, of the
application of a specified set of theoretical assumptions to the actual
outcome of an actual experiment. As Stapp notes, counterfactual
statements of this kind are ``endemic in science'' and the
wholesale rejection of all arguments containing such statements ``is
not rational''.
\qquote{Statements of this general kind are commonplace in physics:
Theory often allows one to deduce from the outcome of certain
measurements on a system what the outcome of some alternative possible
measurements would necessarily be.}
\ni The role of counterfactuals in his argument, then, is this:
\qquote{[T]heoretical assumptions often allow one to say with 
     certainty, on the basis of the outcome of a certain 
     experiment, what ``would have happened'' if an alternative
     possible apparatus had been used.  If logical arguments of 
     this kind -- that follow directly from certain theoretical
     assumptions -- lead to a contradiction, then one must,
     if rational, accept the fact that at least one of the 
     explicit or implicit theoretical assumptions is incorrect.}
\ni The theoretical assumptions to be tested in this way are the
validity of the quantitative predictions of quantum mechanics, Lorentz
invariance, and certain specific conditions of locality.  Contingent
on the validity of quantum mechanics, a contradiction would exclude
Lorentz invariance or locality.

Keeping in mind Stapp's clear, convincing, and explicit
characterization of the proper role for counterfactual conditionals in
physical reasoning, in Section III I point out an essential ambiguity
in the notion of ``present facts'' used to justify the step from the
fifth to the sixth of his fourteen propositions, that undermines the
soundness of his conclusion.  Readers familiar with Stapp's argument
can go directly to Section III.  Others may want to look at Section
II, where I provide the background necessary for the point I wish to
make.  In Section IV I indicate how the nature of this hidden gap in
Stapp's tight and subtle argument might offer some illumination to
those who have difficulty\ft{See for example John S.~Bell,
``Bertlmann's socks and the nature of reality,'' Appendix 1, in {\it
Speakable and Unspeakable in Quantum Mechanics\/}, Cambridge (1987),
155-156.} understanding\ft{For a thoughtful critique of Bell's
statements about Bohr's views, see Arkady Plotnitsky, {\it
Complementarity}, Duke University Press (1994), 172-190.} Bohr's
reply\ft{Niels Bohr, ``Can quantum-mechanical description of physical
reality be considered complete?'', Phys.~Rev.~{\bf 48}, 696-702
(1935).} to Einstein, Podolsky, and Rosen\ft{Albert Einstein, Boris
Podolsky, and Nathan Rosen, ``Can quantum-mechanical description of
physical reality be considered complete?'', Phys.~Rev.~{\bf 47},
777-780 (1935).} (EPR).\ft{From a somewhat different perspective and
within a broader treatment of Stapp's argument, W. Unruh identifies
essentially the same weakness (in his discussion of Stapp's ``LOC2'')
in ``Is quantum mechanics non-local?'', quant-ph 9710032.  My interest
here is in the remarkable way Bohr's critique of EPR is clarified by
applying it to Stapp's argument.}

\bigskip
\centerline{{\bf II.  Background.}}
\medskip 

Stapp's analysis of the data in the Hardy experiment is so cleverly
constructed, so economically stated, and so formally expressed, that
it takes (or, more accurately, it took me) a major effort to follow
it.  Those who make the effort will be rewarded by the beautiful
artistry of the construction.  Because, however, the disabling
ambiguity I wish to identify occurs early in the argument, and because
I am primarily interested here in the light it sheds on Bohr's reply
to EPR, I give here a statement of the Hardy experiment (in Stapp's
notation) that ought to permit a reader to follow the subsequent
discussion of that ambiguity (Section III) and its relation to Bohr's
reply to EPR (Section IV), without having to master the subtleties of
Stapp's full argument or the earlier papers of Hardy. 

Hardy's experiment is of the EPR type: two correlated particles fly
apart to two experimental stations on the left ($L$) and right ($R$).
At each station a property independent of their orbital degrees of
freedom (e.g.~spin or polarization) is subject to one of two
alternative measurements ($L1$ or $L2$ on the left, and $R1$ or $R2$
on the right), each with two possible outcomes (+ or $-$).  The
correlations in the outcomes of the four possible pairs of
measurements are encapsulated in the two-particle state, given (to
within a normalization constant) by $$\k\Psi =
\k{L1+,R1-} -
\k{L2-,R2+}\,\ip{L2-,R2+}{L1+,R1-}.\eq(Psi)$$ Here a state such as
$\k{L1+,R1-}$ indicates a simultaneous eigenstate of the commuting
observables $L1$ and $R1$ with eigenvalue + on the left and $-$ on the
right.  

Note that the state $\k\Psi$ has been constructed to be orthogonal to
the state $\k{L2-,R2+}$:  $$\ip{L2-,R2+}\Psi = 0.\eq(orth22)$$
Furthermore each of the two terms in \(Psi) is easily seen to be
orthogonal to either of the states $\k{L1-,R2-}$ or $\k{L2+,R1+}$:
$$\ip{L1-,R2-}\Psi = 0,\eq(orth12)$$ $$\ip{L2+,R1+}\Psi = 0.
\eq(orth21)$$ Finally, because the observables 1 and 2 are
non-trivially different on both left and right, $\k\Psi$ is not
orthogonal to $\k{L1-,R1+}$:  $$\ip{L1-,R1+}\Psi \neq 0,\eq(north11)$$

The implications of \(orth22), \(orth12), and \(orth21) are,
respectively, that neither 22$-+$, 12$--$, nor 21++, can occur. (By a
symbol like 21+$-$ I mean that experiment 2 is performed on the left
and experiment 1, on the right, with results + on the left and $-$ on
the right.)  The implication of \(north11) is that 11--+ {\it can\/}
occur. Putting this in terms of joint probabilities, $$p(12--) = 0,
\eq(p12)$$ $$p(22-+) = 0, \eq(p22)$$ $$p(21++) = 0,
\eq(p21)$$ $$p(11-+) \neq 0. \eq(p11)$$ These four probabilities are
the only features of the Hardy state needed for an understanding of
what follows.  The consequences of \(p12)-\(p11) that Stapp requires
for his analysis are these: 

\qquote{If experiment 1 is performed on the left and the outcome is
$-$ then if experiment 2 is performed on the right the outcome must be
+ [required by \(p12)].}

\qquote{If experiment 2 is performed on the right and the outcome is
$+$ then if experiment 2 is performed on the left the outcome must be
+ [required by \(p22)].}

\qquote{If experiment 2 is performed on the left and the outcome is
$+$ then if experiment 1 is performed on the right the outcome must be
$-$ [required by \(p21)].}

\qquote{If experiment 1 is performed on the left and the outcome is
$-$ then if experiment 1 is performed on the right the outcome may be
+ [allowed by \(p11)]}
\ni For ease in referring back to them I summarize these four
assertions (listing the compact summaries in the same order) by:

$$L1- \Longrightarrow R2+, \eq(L1-:R2+)$$
$$R2+ \Longrightarrow L2+, \eq(R2+:L2+)$$
$$L2+ \Longrightarrow R1-, \eq(L2+:R1-)$$
$$L1- \Longrightarrow \hskip -0.23 truein / \hskip 0.23 truein R1-. 
\eq(L1-:R1-)$$

Note that a superficial glance the first three of the four assertions
suggest that since $L1+$ implies $R2+$, $R2+$ implies $L2+$, and $L2+$
implies $R1-$, simple logic requires that $L1+$ implies $R1-$, which
contradicts the fourth.  This, of course, is wrong, for these are
assertions about the outcomes of four different possible experiments
(12, 22, 21, and 22) only one of which can actually be done.  At most
one of them can be valid.  It is nonsense to reason in this way from
all four at once, and Stapp does not do this.  

What he does try to do, interestingly, is to set this crude intuition
in a context that allows the logic suggested by the superficial glance
to operate legitimately, even though only one of the four experiments
can actually be done on any given pair of particles.  He gains the
necessary freedom to maneuver by considering the case in which the
decisions of which measurement to make on both left and right, along
with the performance and recording of the outcomes of those
measurements, are confined to space-like separated regions of
space-time.  Lorentz invariance then requires that any facts, valid in
a frame of reference in which the choice and outcome of an experiment
on one side happens before the choice and outcome on the other side,
must have a valid formulation in a frame in which the time order is
reversed.  The goal of this strategy is to reduce all statements about
the measurements that might have been made but were not, to deductions
that exploit the general principle (locality) that decisions about
which measurement to make on one side, cannot alter the factual status
of an {\it earlier\/} choice, performance, and outcome of a measurement on
the other side.   If one could achieve this goal, then the resulting
contradiction would justify Stapp's claim, cited above, about present
facts depending on future choices. 

\bigskip
\centerline{{\bf III. What's wrong}} 
\medskip

Because the choice, performance, and outcome of the measurement on one
side can be regarded as taking place either before or after the
space-like-separated choice, performance, and outcome of the
measurement on the other side, all statements about any of these
events must have a valid formulation from either point of view, or in
frame-independent atemporal terms.  For example the feature
summarized by \(L1-:R2+) can be stated atemporally as 

\qquote{If 1 is measured on the left with the result $-$ then if 2 is
measured on the right the result must be +.}
\ni Or it can be stated as 
\qquote{If 1 was measured on the left with the result $-$ then if 2 is
subsequently measured on the right the result must be +.}
\ni Or it can be stated as 
\qquote{If 1 is measured on the left with the result $-$ then if 2 was
earlier measured on the right the result must have been +.}

Bearing this in mind, consider Stapp's fifth proposition, which
translated out of his formal notation into everyday language states
(in atemporal terms) that 
\qquote{(I)\ \ Whenever the choice of measurement on the left is $L2$, if the
measurement on the right is $R2$ and gives +, then if $R1$ were instead
performed the result would be $-$.} 
\ni (The ``were'' here is the ``were'' of the present subjunctive ---
not the past indicative.)  The validity of this is established by
translating it into the language appropriate to the frame of reference
in which the events on the left happen first:
\qquote{(I$_L$)\ \ Whenever the choice of measurement on the left was
$L2$, if the measurement done later on the right is $R2$ and gives +,
then if $R1$ were instead done later on the right, the result would
have to be $-$.} 
\ni To see that (I$_L$) is correct, note that it states that $L2$ was
performed on the left, but does not specify the outcome.  If
subsequently $R2$ is performed on the right and gives +, this enables
us to deduce from
\(R2+:L2+) that the result of the earlier measurement $L2$ must have
been +.  This in turn enables us to deduce from
\(L2+:R1-) that if the choice at $R$ had been to measure 1, the result
would certainly have been $-$.

The use here of a counterfactual conditional is entirely in accord
with Stapp's criterion for its legitimacy.  It is based on the outcome
of an actual experiment (the measurement of $L2$ on the left with the
outcome +) and a theoretical assumption (the validity of the quantum
mechanical calculation that the joint probability of $L2+$ and $R1-$
in the Hardy state is zero.)

Note next that Lorentz invariance requires (I) to be valid when stated
in a language appropriate to a frame in which the events on the right
happen first:  
\qquote{(I$_R$)\ \ Whenever the choice of measurement on the left is
$L2$, if the measurement done earlier on the right was $R2$ and gave
+, then if $R1$ had instead been done earlier on the right, the result
would necessarily have been $-$.}  \ni The only thing preventing one
from establishing (I$_R$) directly by essentially the same argument
that established (I$_L$), is the objection that there are no compelling
grounds for assuming that the result of the later experiment on the
left in the hypothetical case ($R1$ performed earlier on the right)
should be the same as the actual result in the actual case ($R2$
performed earlier on the right).  This objection is blocked, however,
by the fact that it makes no sense in the frame in which the
experiment on the left happens first, and that the recorded result of
an actual experiment is a Lorentz invariant piece of information.

Now comes the crucial step to Stapp's sixth proposition [called here
(II)].  That proposition is the same as the fifth [(I) above] except
that $L2$ has been changed to $L1$:
\qquote{(II) Whenever the choice of measurement on the left is $L1$, if the
measurement on the right is $R2$ and gives +, then if $R1$ were instead
performed the result would be $-$.} 
\ni If we can legitimately get to (II), then the rest is almost a
mopping up operation.  Informally, suppose the result of that $L1$
measurement happens to be $-$ (which \(p11) ensures has a non-zero
probability).  Then \(L1-:R2+) guarantees that the result of the
measurement of $R2$ on the right must be +, and (II) then tells us
that if $R1$ were instead performed on the right the result would be
$-$.  The desired contradiction comes from
\(L1-:R1-), which tells us that the result $L1-$ need not necessarily
be accompanied by $R1-$.  There is, to be sure, some tricky terrain to
traverse in making these further informal steps airtight, but Stapp
brings considerable ingenuity to bear on that task. I do not explore
here whether there are further pitfalls, because the move from (I) to
(II) is already unsupportable, and for interesting reasons. 

Stapp's justification for the move from (I) to (II) is based on viewing
(I) in the language of the frame in which the events on the right
happen first --- i.e. taking (I) in the form (I$_R$) given above.
Consider now the phrase appearing in $(I_R)$,
\qquote{(S) if the measurement done earlier on the right was $R2$ and
gave +, then if $R1$ had instead been done earlier on the right, the
result would necessarily have been $-$.}
\ni Stapp asserts that this refers only to the state of affairs on the
right at times prior to the choice of what measurement to make on the
left.  Since everything of interest on the right takes place before
the choice and execution of the experiment on the left, and since
``the truth of a statement that refers only to phenomena confined to
an earlier time cannot depend upon which measurement an experimenter
will freely choose to perform at a later time'', it is
permissible to replace $L2$ with $L1$.  

The problem with this justification for the replacement of $L2$ with
$L1$ is that it goes beyond Stapp's criterion for the proper use of
counterfactual conditionals.  The first part of (S), \qquote{(S$_1$)
$R2$ was performed earlier and gave +,} \ni does refer to an actual
event on the right prior to the choice of experiment on the left, and
it does indeed make no sense to claim that its meaning or validity
depends on that subsequent choice.  But the rest of the statement (S),
\qquote{(S$_2$) if $R1$ had been performed earlier instead it would
have given $-$,} \ni is quite another matter.  This is a
counterfactual conditional.  Unlike the meaning and validity of
(S$_1$), the meaning and validity of ($S_2$) do not derive directly
from an actual performance of $R1$ giving the actual result $-$.  As a
counterfactual conditional it only has meaning as an inference from
actual results of actual experiments in combination with theoretical
principles.  One piece of information going into that inference is
what actually happened on the right ($R2$ was performed and gave the
result +).  So far so good.  But the other fact, required to make that
inference, is that the subsequent choice of experiment on the left
turns out, in fact, to be $L2$.  You cannot make the inference
asserted by the counterfactual conditional if the subsequent choice of
experiment on the left turns out to be $L1$.  {\it A statement about
the outcome of an experiment at a given time that might have been
performed but in fact was not, may derive its truth or falsity and,
indeed, its very meaning from events in the future that have not yet
happened.} The statement $(S_2)$ and hence the full statement $(S)$
implicitly does refer to a choice taking place on the left at the
later time.

Overlooking this difficulty in the passage from (I) to (II), Stapp
takes his additional steps to the final contradiction, and concludes
that at least in some frames of reference the truth of a statement
referring only to phenomena confined to an earlier time can depend on
which measurement an experimenter freely chooses to perform at a later
time.  This does not mean, however, that a choice made in the future
(whether to measure $L1$ or $L2$ on the left) can influence events in
the present.  What it does mean is that {\it a choice made in the
future can influence statements about events in the present that might
have happened but did not.\/} It is this that prohibits the step from
(I) to (II).  The choice of experiment on the left cannot affect {\it
what actually happened\/} in the past on the right.  But it can affect
{\it the kinds of inferences\/} one can make about hypothetical
behavior on the right.  Such inferences are required to give meaning
to a counterfactual conditional according to Stapp's own rule. 
  
\bigskip
\centerline{{\bf IV. Bohr's Reply to EPR}}
\medskip
\nobreak
This refutation of Stapp's nonlocality argument is strikingly evocative
of Bohr's reply to EPR: 
\qquote{[T]here is $\ldots$ no question of a mechanical disturbance of
   the system under investigation $\ldots$ [but] there is essentially
   the question of an influence on the very conditions which define
   the possible types of predictions regarding the future behavior of
   the system.}  \ni This fragment of his longer article is virtually
the entire point of his earlier very brief response,\ft{Niels Bohr,
``Quantum mechanics and physical reality'', Nature {\bf 136}, 65
(1935)} where he puts it this way: \qquote{It is true that in the
measurements under consideration any direct mechanical interaction of
the system and the measuring agencies is excluded, but a closer
examination reveals that the procedure of measurements has an
essential influence on the conditions on which the very definition of
the physical quantities in question rests.}

In the case of Stapp's argument there is no possibility of the
subsequent choice of measurement on the left having an effect on the
earlier choice, performance, or outcome of the measurement on the
right.  But that choice on the left does have an influence on the
condition that defines the very meaning of counterfactual statements
about what might have happened earlier on the right.  Since there are
no grounds for imposing on counterfactual statements a requirement
that they derive their meaning solely from conditions at or prior
to the time to which they refer, we see that Stapp's argumentation
does not justify his conclusion that ``phenomena confined to an
earlier time'' can in some Lorentz frame depend upon which measurement
an experimenter will freely choose to perform at a later time.  

The illustration of Bohr's remark through Stapp's argument makes clear
that in denying the existence of a ``mechanical disturbance'' while
maintaining the existence of an ``influence'' Bohr is in no way
asserting the presence of a mysterious non-mechanical disturbance
(``quantum nonlocality'').  In Stapp's analysis of the Hardy
experiment, the influence of the choice of what to measure on the left
only is only an influence on whether or not a counterfactual statement
about events happening entirely on the right has any meaning.  Stapp's
argument does not demonstrate nonlocality because that choice of what
to measure on the left alters {\it no thing\/} on the right --- i.e.
it is ``not a mechanical disturbance.'' What it does alter is {\it
what we can meaningfully say\/} about events on the right --- i.e. it
is an influence on the conditions that permit us to make a meaningful
counterfactual statement.

In comparing this analysis of Stapp to Bohr's analysis of EPR one
encounters an interesting difference in the time order of
corresponding events.  (Stapp, of course, uses both time orders, but
the justification offered for the crucial replacement of $L2$ with
$L1$ is argued in the frame in which the event on the right happens
first.)  In EPR the choice of what to measure on one side influences
not the ability to talk counterfactually about a past measurement on
the other side, but the ability to make a prediction about a future
measurement on the other side.  The prediction that EPR lose the
ability to make as a result of making a different past choice,
corresponds to the past counterfactual Stapp loses the ability to give
meaning as a result of making a different future choice.  It is easier
to grasp the point corresponding to Bohr's in the context of Stapp's
argument than in the context of EPR, because we are used to thinking
of counterfactuals as mere statements not necessarily about anything
real.  Predictions with certainty on the other hand --- even those
that our present actions prevent us from being able to make --- we are
tempted, with Einstein, Podolsky, and Rosen, to view as statements
about elements of reality.

In Stapp's case a different choice of what to measure on one side has
an influence on the very conditions which define the possible types of
valid counterfactual statements regarding the past behavior of the
system.  In the EPR case a different choice of what to measure on one
side has ``an influence on the very conditions which define the
possible types of predictions regarding the future behavior of the
system'' on the other side.  In both cases the influence is not on a
far away actual state of affairs, but on our ability to make valid
statements about that far away state of affairs.  A counterfactual is
validated by two things: a past event (the actual measurement whose
result is required to confirm the counterfactual) and the right future
choice of measurement (that makes possible the theoretical
deduction).  A prediction is validated by two things: a future event
(the actual measurement whose result is required to confirm the
prediction) and the right past choice (that makes possible the
prediction).  

By bringing us full circle from Bell and Hardy all the way back to EPR
and Bohr, but with an interesting twist, Stapp has provided us with a
wonderful opportunity to understand better and to savour more fully,
the extraordinary content of that profoundly subtle exchange. 

\bigskip

{\sl Acknowledgment.} This analysis was supported by the National Science Foundation under
Grant No.~PHY9722065.

\bye